# Modeling *In vivo* Wireless Path Loss

Yang Liu, Thomas P. Ketterl, Gabriel E. Arrobo, and Richard D. Gitlin

Department of Electrical Engineering
University of South Florida, Tampa, Florida 33620, USA
Email: {yangl, garrobo}@mail.usf.edu, {ketterl, richgitlin}@usf.edu

*Abstract* — Our long-term research goal is to model the *in vivo* wireless channel. As a first step towards this goal, in this paper we performed *in vivo* path loss measurements at 2.4GHz and make a comparison with free space path loss. We calculate the path loss by using the electric field radiated by a Hertzian-Dipole located inside the abdominal cavity. The simulations quantify and confirm that the path loss falls more rapidly inside the body than outside the body. We also observe fluctuations of the path loss caused by the inhomogeneity of the human body. In comparison with the path loss measured with monopole antennas, we conclude that the significant variations in Received Signal Strength is caused by both the angular dependent path loss and the significantly modified *in vivo* antenna effects.

## I. INTRODUCTION

The wireless body area network (WBAN) [1] IEEE 802.15 Task Group 6 studied the devices and technologies on, in or around the human body for various kinds of applications such as healthcare and entertainment. However, research on *in vivo* models for propagation in the human body is still in the early stages. The characteristics of the *in vivo* channel are significantly different than those of classical wireless cellular and Wi-Fi systems. Understanding the characteristics of the *in vivo* channel is necessary to optimize *in vivo* physical layer signal processing, and designing efficient networking protocols that ultimately will make possible the deployment of wireless body area networks inside the human body.

There are many challenges in characterizing the *in vivo* channel including the inhomogeneous and very lossy nature. Furthermore, additional factors need to be taken into account, such as near-field effects and highly variable propagation speeds through different organs and tissues. These effects are summarized in Table I and illustrated in Fig. 1.

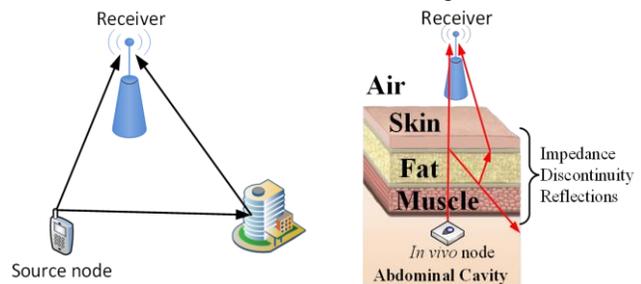

Fig. 1. Classic multi-path channel vs *in vivo* multi-path channel

In this paper, we study the path loss for *in vivo* wireless communications. The rest of the paper is organized as follows. In section II, we summarize the prior work on *in vivo* wireless communications and channel modelling in WBANs. Our simulation setup and the approach to obtain the path loss for the *in vivo* channel are described in section III. In Section IV, our simulation results and analysis for *in vivo* path loss are presented. Finally, in Section V we present our conclusions and future research directions.

TABLE I
COMPARISON OF *EX VIVO* AND *IN VIVO* CHANNEL

| Feature | *Ex vivo* | *In vivo* |
|---|---|---|
| Physical Wave Propagation | Constant speed<br>Multipath − reflection, scattering and diffraction | Variable speed<br>Multipath − plus penetration |
| Attenuation and Path Loss | Lossless medium<br>Decreases inversely with distance | Very lossy medium<br>Angular (directional) dependent |
| Dispersion | Multipath delays → time dispersion | Multipath delays of variable speed → frequency dependency → time dispersion |
| Directionality | Propagation essentially uniform | Propagation varies with direction<br>Directionality of antennas changes with position/orientation |
| Near Field Communications | Deterministic near-field region around the antenna | Inhomogeneous medium → near field region changes with angles and position inside body |
| Power Limitations | Average and Peak | Plus specific absorption rate (SAR) |
| Shadowing | Follows a *log-normal* distribution | To be determined |
| Multipath Fading | Flat fading and frequency selective fading | To be determined |
| Antenna Gains | Constant | Angular and positional dependent<br>Gains highly attenuated |
| Wavelength | The speed of light in free space divided by frequency | $\lambda = \frac{c}{\sqrt{\varepsilon_r}f}$ → at 2.4GHz, average dielectric constant $\varepsilon_r = 35$ → roughly 6 times smaller than the wavelength in free space. |

## II. LITERATURE REVIEW

### A. In vivo Wireless Communications

Understanding the *in vivo* wireless channel is critical to advancing many bio-medical and other procedures. The authors [2] performed signal strength and channel impulse response simulations using an accurate human body model and investigated the variation in signal loss at different RF frequencies as a function of position around the human body. In [3], the maximum allowable transmitted power levels for *in vivo* devices was studied in order to achieve a required bit error rates (BER) at the external node (receiver) while maintaining the specific absorption rate (SAR) under a required threshold. However, the previous research does not include the fundamental characterization of the *in vivo* channels, is the focus of this paper.

### B. In vivo Channel Characterization

For *in vivo* channel modeling, a phantom or a human body model is necessary to be used for measurement. For example, in [4], the authors observed the radio frequency (RF) propagation from medical implants inside a human body via a 3D Immersive Platform. An *in vivo* channel model for homogeneous human tissues was developed in [5]. Using ingested wireless implants, the authors in [6] performed numerical and experimental investigations for biotelemetry radio channels and wave attenuation in human subjects.

## III. SIMULATION SETUP

### A. Human Body Model

We use the ANSYS HFSS 15.0.3 Human Body Model software to perform our simulations. This tool contains an adult male body with more than 300 parts of muscles, bones and organs modeled to 1 mm. The antenna we use is the Hertzian-Dipole, which can be treated as an ideal dipole. In this way, we can investigate the path loss when there is little antenna effect. It will help to explore the effects of using different types of antennas. The operating frequency is the 2.4 GHz ISM band.

### B. Measurement approach

Since the *in vivo* environment is an inhomogeneous medium, it is instructive to measure the path loss in the spherical coordinate system. The truncated human body, the Hertzian-Dipole and the spherical coordinate system are shown in Fig. 2.

The path loss can be calculated as:

$$Path\ loss(r,\theta,\phi) = 10 * \log_{10}\left(\frac{|E|^2_{r=0}}{|E|^2_{r,\theta,\phi}}\right) \quad (1)$$

where $r$ represents the distance from the origin, i.e. the radius in spherical coordinates, $\theta$ is the polar angle and $\phi$ is the azimuth angle. $|E|^2_{r,\theta,\phi}$ is the square of the magnitude of the electric $E$ field at the measuring point and $|E|^2_{r=0}$ is the square of the magnitude of $E$ field at the origin.

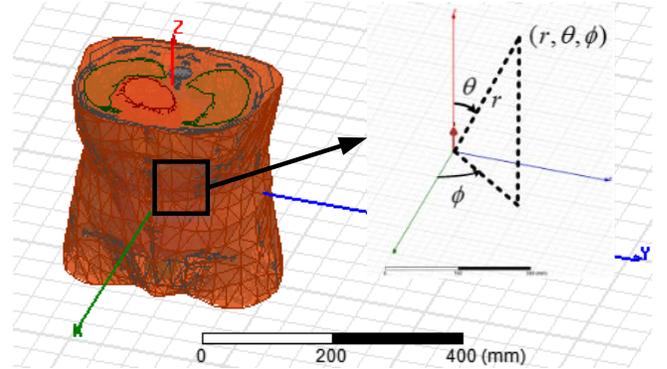

Fig. 2. Truncated human body with Hertzian-Dipole at the origin in spherical coordinate system

## IV. RESULTS

### A. Path loss vs distance

When we fix the azimuth and polar angles to 0° and 90°, respectively, we obtain the relationship between path loss and distance, as shown in Fig. 3. For the *in vivo* case, the skin boundary is at $r = 108mm$. We can clearly observe the different behavior of the path loss between the *in vivo* and *ex vivo* regions. In the body, the path loss drops rapidly and there exist some standing waves that are caused by the impedance mismatch between the two media. Outside the body, the path loss tends to fall more smoothly.

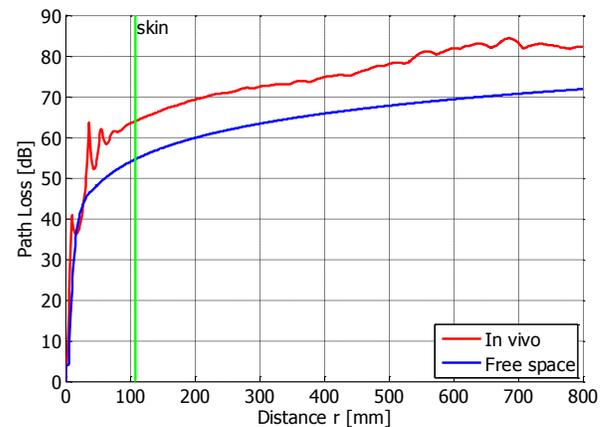

Fig. 3. Path loss vs. distance at azimuth angle $\phi = 0°$ and polar angle $\theta = 90°$

In contrast, in the range of $r = 70 \sim 400\ mm$, the *in vivo* path loss is about 9 dB greater than the free space path loss. Both the free space and *in vivo* path losses initially fall rapidly, but the *in vivo* path loss falls rapidly inside the body while free space path loss also does so for $r = 1 - 20\ mm$, which is exactly the near field region.

## B. Path loss vs azimuth angle

In this simulation, we vary the distance $r = 300\ mm/200mm/100mm$ and fix the polar angle at $\theta = 90°$. In this way, we obtain the path loss vs azimuth angle as shown in Fig. 4. The curve of *in vivo* path loss at $r = 100mm$ almost overlaps with the one of free space path loss at $r = 300mm$. Overall, the *in vivo* path loss is about 9 dB greater than the free space path loss. We can see that the free space path loss is flat and the *in vivo* path loss has small fluctuations up to 1dB. These fluctuations show that the human body is only mildly inhomogeneous and, consequently, that the path loss is slightly angular dependent. In our previous research [2], we found that significant variations in Receive Signal Strength (RSS) could be observed with only very slight variations of angular position (by as much as 20 dB) when a monopole antenna was used. In the measurements in this paper, we use the Hertzian-Dipole, which minimizes the antenna effects on the path loss.

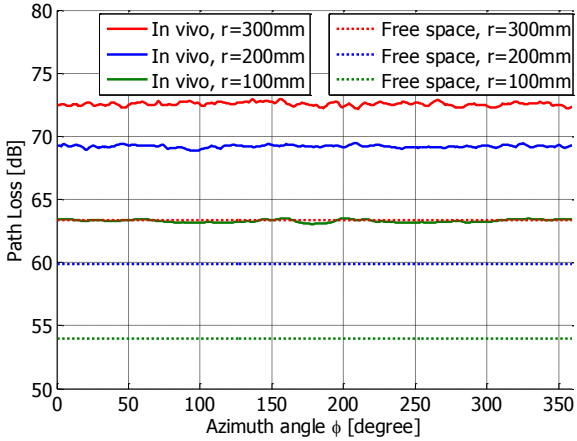

Fig. 4. Path loss vs azimuth angle at polar angle $\theta = 90°$ and distance $r = 300mm, 200mm, 100mm$

## C. Path loss vs polar angle

Figure 5 shows the path loss vs polar angle when the distance $r = 150mm/100mm/50mm$ and azimuth angle $\phi = 0°$. For the cases of $r = 150mm$ and $r = 100mm$, the *in vivo* path loss curve almost has the same shape as the one for free space and it is about 9dB greater than the free space path loss. The reason why the curve of the free space path loss appears as an arch instead of a flat line is that the Hertzian-Dipole has some effects on the path loss in different polar angles because of its donut-shaped antenna pattern. However, when the distance $r = 50mm$, the free space path loss is almost a flat line, which means that there is little antenna effect. For the *in vivo* path loss at $r = 50mm$, we observe that there are two arches at $\theta = 25° - 50°\ and\ 60° - 130°$. This is because the path is passing through the small intestine, which makes the path loss relatively greater.

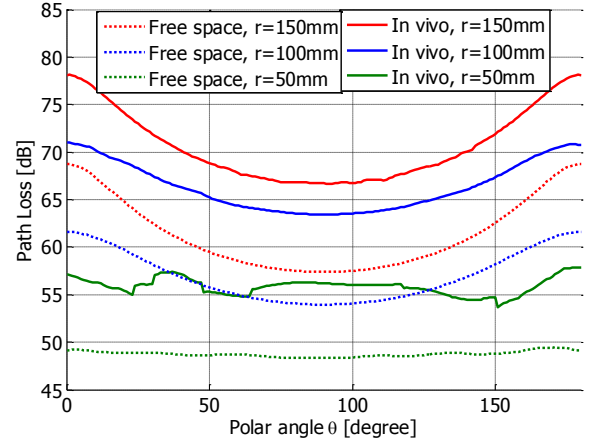

Fig. 5. Path loss vs polar angle at azimuth angle $\phi = 0°$ and distance $r = 150mm, 100mm, 50mm$

## V. CONCLUSION AND FUTURE RESEARCH

We used HFSS software and Human Body Model to calculate the electric field caused by a Hertzian-Dipole at the origin and obtained the *in vivo* path loss versus different parameters in spherical coordinates. From our initial results we observed the different behaviors of the path loss between *in vivo* and *ex vivo* environments. Outside of the body, the *in vivo* path loss has small fluctuations and is 9dB greater than the free space path loss. Inside the body, the *in vivo* path loss has some standing waves and is also impacted by the organs. We also compared the results to the method of using monopole antennas and found the angular dependent signal variation was caused by both the angular based path loss and *in vivo* antenna effects.

This initial research is a first-step in building an *in vivo* channel model and in exploring the different types of *in vivo* antenna effects.

## ACKNOWLEDGEMENT

This publication was made possible by NPRP grant # 6-415-3-111 from the Qatar National Research Fund (a member of Qatar Foundation). The statements made herein are solely the responsibility of the authors.